\documentclass[useAMS]{mn2e}
\usepackage{graphicx}
\usepackage{amsfonts}
\pdfoutput=1 
\title[]{Segregation Effects According to the Evolutionary Stage of Galaxy Groups}
\author[A.L.B. Ribeiro, P.A.A. Lopes and M. Trevisan]
{A.L.B. Ribeiro$^{1}$\thanks{E-mail: albr@uesc.br} P.A.A. Lopes$^{2}$ and M. Trevisan$^{3}$ \\
$^{1}$ Laborat\'orio de Astrof\'{\i}sica Te\'orica e Observacional, Universidade Estadual de Santa Cruz -- 45650-000, Ilh\'eus-BA, Brazil\\
$^{2}$ Observat\'orio do Valongo, Universidade Federal do Rio de Janeiro, Brazil\\
$^{3}$ Instituto Astron\^omico e Geof\'{\i}sico- USP, S\~ao Paulo-SP, Brazil}
\begin{document}

\date{Accepted 2010 September 20. Received 2010 July 19}

\pagerange{\pageref{firstpage}--\pageref{lastpage}} \pubyear{2010}

\maketitle

\label{firstpage}

\begin{abstract}
We study segregation phenomena in 57 groups selected 
from the 2PIGG catalog of galaxy groups. The sample corresponds to those systems
located in areas of at least 80\% redshift coverage out to 10 times the 
radius of the groups. The dynamical state of the galaxy systems was determined after
studying their velocity distributions. We have used the Anderson-Darling test
to distinguish relaxed and non-relaxed systems. This analysis indicates that
84\% of groups have galaxy velocities consistent with the normal distribution, while
16\% of them have more complex underlying distributions. Properties of the member galaxies 
are investigated taking into account this classification. 
Our results indicate that galaxies in Gaussian groups are significantly more evolved than galaxies in non-relaxed systems out to distances of $\sim 4R_{200}$, presenting significantly redder (B-R) colors. We also find evidence that galaxies with $M_R \le -21.5$ in Gaussian groups are closer to the condition of energy equipartition.

\end{abstract}


\section{Introduction}
Groups of galaxies contain about half of all galaxies in the
Universe (e.g., Huchra \& Geller 1982; Geller \&
Huchra 1983; Nolthenius \& White 1987; Ramella et al. 1989). 
They represent the link between galaxies and large-scale structures
and play an important role to galaxy formation and evolution.
One of the most important questions about galaxy systems is related to
segregation phenomena. The study of segregation effects is important
to understand how system environment is transforming galaxies at the
present epoch. Evidence for different loci in position and velocity
spaces according to luminosity, spectral type and color of galaxies suggests ongoing evolution
of clusters through the process of mergers, dynamical friction and secondary infall
(e.g. Adami, Biviano \& Mazure 1998, Biviano et al. 2002). Segregation has also been observed
in galaxy groups (e.g. Mahdavi et al. 1999, Carlberg et al. 2001), suggesting a
continuum of segregation properties of galaxies from low-to-high mass systems
(Girardi et al. 2003). However, the dynamical state of galaxy groups
is not taken into account in these studies. Differences in segregation phenomena 
may emerge if one divides groups according to their evolutionary stage. Recently, Hou et al. (2009) have examined three goodness-of-fit tests (Anderson-Darling, Kolmogorov and $\chi^2$ tests) to find which statistical tool is best able to distinguish between relaxed and non-relaxed galaxy groups. Using Monte Carlo simulations and
a sample of groups selected from the CNOC2, they found that the Anderson-Darling (AD) test
is far more reliable at detecting real departures from normality. Their results show
that Gaussian and non-Gaussian groups present distinct velocity dispersion profiles, suggesting that discrimination
of groups according to their velocity distributions may be a promising way to access 
the dynamics of galaxy systems. Extending up this kind of analysis to the outermost edge of groups
one can probe the regions where they might not be in dynamical equilibrium.
In this letter, we look for segregation effects  in galaxy groups selected from the 2PIGG catalog (Eke et al. 2004), using 2dF data out to 4$R_{200}$, and taking into account the evolutionary stage of the groups resulting from the AD test.

\section{Data and Methodology}

\subsection{2PIGG sample}

We use a subset of the 2PIGG catalog,
corresponding to groups located in areas
of at least 80\% redshift coverage in 2dF data out to 10 times the
radius of the systems, roughly estimated from the projected harmonic mean (Eke et al. 2004).
The idea of working with such large areas is to probe the effect of
secondary infall onto groups. Members and interlopers were redefined after the
identification of gaps in the redshift distribution according
to the technique described by Lopes et al. (2009). 
Before selecting group members and rejecting interlopers we first
refine the spectroscopic redshift of each group and identify its
velocity limits. For this purpose, we employ the gap-technique described 
in Katgert et al. (1996) and Olsen et al. (2005) to identify 
gaps in the redshift distribution. A variable gap, called 
{\it density gap} (Adami et al. 1998), is
considered. To determine the group redshift, only galaxies within 
0.50 h$^{-1}$ Mpc are considered. Details about this procedure are found 
in Lopes et al. (2009); see also Ribeiro et al. (2009) for applications
of this technique to 2dF galaxy groups. With the new redshift and 
velocity limits, we apply an algorithm for interloper 
rejection to define the final list of group members. We use the
``shifting gapper'' technique (Fadda et al. 1996), which consists of the 
application of the gap-technique to radial bins from the group center.
We consider a bin size of 0.42 h$^{-1}$ Mpc (0.60 Mpc for h = 0.7) or 
larger to ensure that at least 15 galaxies are selected. Galaxies not 
associated with the main body of the group are discarded. This procedure 
is repeated until the number of group members is stable and no further galaxies 
are eliminated as intruders. 
In the present work,
we have sampled galaxies out to 10 times the hamonic mean radius of the systems, including
galaxies whose distances to the centers can reach $\sim$8 Mpc. To avoid contamination
of nearby structures, we select galaxies within the maximum radius $R_{max}=$ 4.0 Mpc
(see La Barbera et al. 2010). After applying the shifting gapper procedure we have a list of group members and we call $R_A$ the aperture equivalent to the radial offset of the most distant member (normally close to $R_{max}$). We estimate the velocity dispersion ($\sigma$) within $R_A$ and then the physical radius (R$_{200}$) of each group.
Finally, a virial 
analysis is perfomed for mass estimation (M$_{200}$). Further 
details regarding the interloper removal and estimation of global properties 
($\sigma$, physical radius and mass) are found in Lopes et al. (2009). 

\subsection{Classifying groups}

The first step in our analysis is to apply the AD test (see Hou et al. 2009 for
a good description of the test) to the velocity
distributions of galaxies in groups. This is done for different distances, producing the following
ratios of non-Gaussian groups: 6\% ($R \leq 1R_{200}$), 9\% ($R\leq 2R_{200}$), 
and 16\% ($R\leq 3R_{200}$ and $R\leq 4R_{200}$). 
Approximately 90\% of all galaxies
in our sample  have distances $\leq 4R_{200}$. This is the natural cutoff in space we have made in this work. Some properties of galaxy groups are presented in Table 1. We have classified groups according to the AD test (at 0.05 significance level) 
done at $R\leq 4R_{200}$ , encompassing all groups with evidence for normality deviations. 
Properties of non-Gaussian (NG) groups in Table 1 were computed twice, with and without
a correction based on iterative removal
of galaxies whose absence in the sample cause the groups become Gaussian, following a procedure
similar to Perea, del Olmo \& Moles (1990). The corrected properties are just those the system would have if it was made only with galaxies
consistent with the normal velocity distribution. This correction allows one to honestly compare typical
properties of G and NG groups. Not doing that, NG groups could have their properties overestimated by a factor of $\sim 1.5$, taking 
all members within $4R_{200}$.  After this procedure, we see in Table 1 that G and corrected NG  groups 
have similar properties.

\subsection{Composite groups}

A suitable way to investigate galaxies in multiple galaxy systems is to combine them
in stacked objects (Biviano et al. 1992). Thus, we built two composite groups, Gaussian--G (composed of 48 systems) and non-Gaussian--NG (composed of 9 systems). Galaxies in theses composite groups have distances to group centers normalized by $R_{200}$ and their velocities refers to the group median velocities and are scaled by the group velocity dispersions

\begin{equation}
u_i={{v_i - \langle v \rangle_j}\over \sigma_j}
\end{equation}

\noindent where $i$ and $j$ are, respectively, the galaxy and the group indices. Velocity 
dispersions of the composite groups refer to the dimensionless quantity $u_i$.
Absolute magnitudes, $M_R$, are obtained from Super-COSMOS R band, a 2dF photometric information.
Cosmology is defined by $\Omega_m$ = 0.3, $\Omega_\lambda$ = 0.7, and $H_0 = 100~h~{\rm km~s^{-1}Mpc^{-1}}$
Distance-dependent quantities are calculated using $h=0.7$. All figures presented in the next section correspond to cumulative data in $R/R_{200}$ or $M_R$.
Error-bars in our analysis are obtained from a bootstrap technique with 1000 resamplings.

\begin{table}
\caption{Mean properties of groups}           
\label{tab1}      
\tiny{              
\begin{tabular}{l c c c c c}       
\hline\hline   
Type          & $R_{200}$ (Mpc) & $M_{200}$ ($10^{14}~M_{\sun}$)& $\sigma $ (km/s) & $N_{200}$ & $N_T$ \\ 
\hline 
G              & $0.94\pm 0.31$  & $0.88\pm 0.79$              & $223 \pm 89$    & $10\pm 4$ & $24\pm 11$ \\
NG             & $1.32\pm 0.27$  & $1.41\pm 0.83$              & $363 \pm 99$    & $12\pm 5$ & $40\pm 12$ \\   
${\rm NG_{c}}$ & $0.97\pm 0.23$  & $0.95\pm 0.95$              & $257\pm 76$     & $10\pm 4$ & $31\pm 10$ \\
\hline
\end{tabular}
}
\end{table}

\section{Segregation Analysis}

Segregation analysis is a powerful tool to evaluate galaxy evolution in 
galaxy systems (e.g. Goto, 2005).
We probe segregation phenomena out to $4R_{200}$, looking for differences in galaxies
with respect to the dynamical state of the groups.
First, we test the presence of luminosity segregation in the velocity space
by computing the normalized velocity dispersion, $\sigma_u$, of the stacked G and NG groups.
In Figure 1, we plot $\sigma_u$ of the composite groups as a function of the absolute magnitude in the R band. We clearly see that, at $M_R \leq -21.5$, the velocity dispersions decreases
towards brighter absolute magnitudes. On the other hand, for fainter absolute magnitudes, 
the velocity dispersions are approximately constant. More interestingly, although the 
result is similar for both stacked groups, for the NG group we see a steeper correlation
in the bright end than that we see for the G group. If one assumes a constant galaxy mass-to-light ratio, energy equipartition implies $\sigma_u \propto 10^{0.2M_R}$ (e.g. Adami, Biviano \& Mazure, 1998). The regression lines between $\log{\sigma_u}$ and $M_R$ have slopes  0.18$\pm$0.05 and 0.38$\pm$0.03, for G and NG groups, respectively. That is, the brightest galaxies are moving more slowly than other group galaxies. Such a segregation in the velocity space may be interpreted as evidence that these galaxies have reached energy equipartition, as a consequence of dynamical friction (e.g Capelato et al. 1981). In fact, the slope we found for galaxies in the G group is consistent with this interpretation. 
However, the steeper relation between $\sigma_u$ and
$M_R$ probably indicates a  departure from equipartition state for galaxies in the NG group.
We also should note that, for $M_R\geq -21.5$, velocity dispersions are larger in the NG group.
Therefore, although fainter galaxies both in G and NG groups  seem to lie in the
velocity equipartition state generated by violent relaxation, 
these galaxies in the NG group have more kinetic energy.
A complementary view of this scenario follows from what is seen in Figure 2. Note that
the velocity dispersion profiles show declining and rising trends, for G and NG groups, respectively. 
They approximately cross each other at $2.5R_{200}$ and then separate more and more
for larger radii. This is consistent with the results of Hou et al. (2009), for the CNOC2
galaxy groups sample. Rising profiles are generally interpreted as a possible signature
of mergers (Menci \& Fusco-Femiano, 1996), which suggests a current intense phase of environmental influence 
on galaxies in the  inner parts of non-Gaussian groups.
Looking  for a counterpart of these effects in color, 
we plot in Figure 3 the color profiles for the G and NG groups. They clearly reveal a stronger reddenning towards the center for galaxies
in the G group. Also, note that the profiles turn flat approximately at 3$R_{200}$, but
galaxies are still redder in Gaussian groups out to 4$R_{200}$. This result indicates that non-Gaussian groups contain less evolved galaxies at the present epoch even in the outskirts. In fact, galaxies in the NG group are fainter than those
in the G group for all radii, with luminosities presenting rising
profiles in both cases (see Figure 4). Spearman tests indicate significant increasing trends up to 1$R_{200}$ and
2.3$R_{200}$ for the G and NG stacked systems, respectively.

\begin{figure}
\includegraphics[width=84mm]{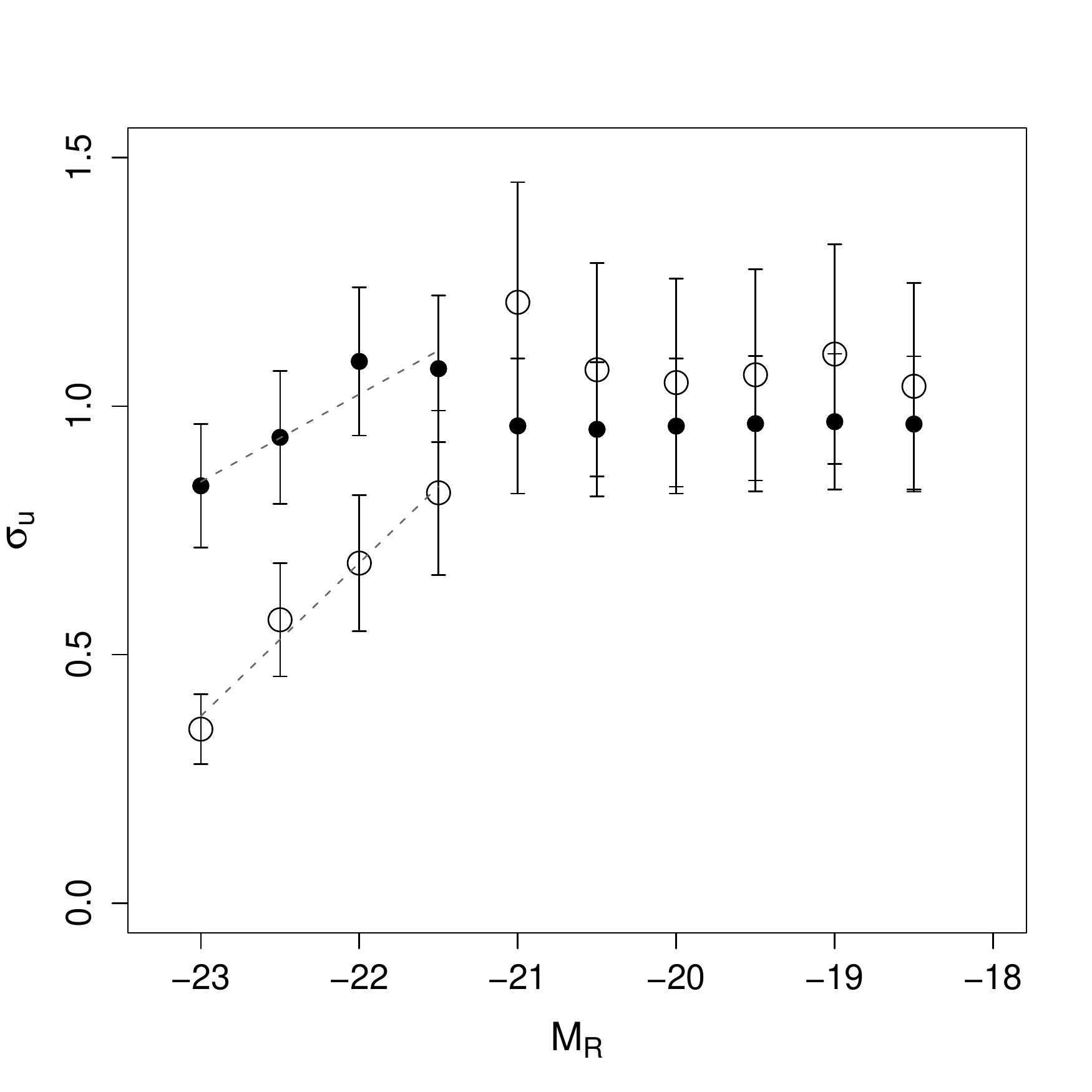}
\caption{Composite groups velocity dispersion as a function of the absolute magnitude in the R band. Filled circles denote galaxies in G groups, while open circles denote galaxies in NG groups. Dashed lines indicate the regression fits for galaxies with $M_R\leq -21.5$.}
\label{}
\end{figure}

\begin{figure}
\includegraphics[width=84mm]{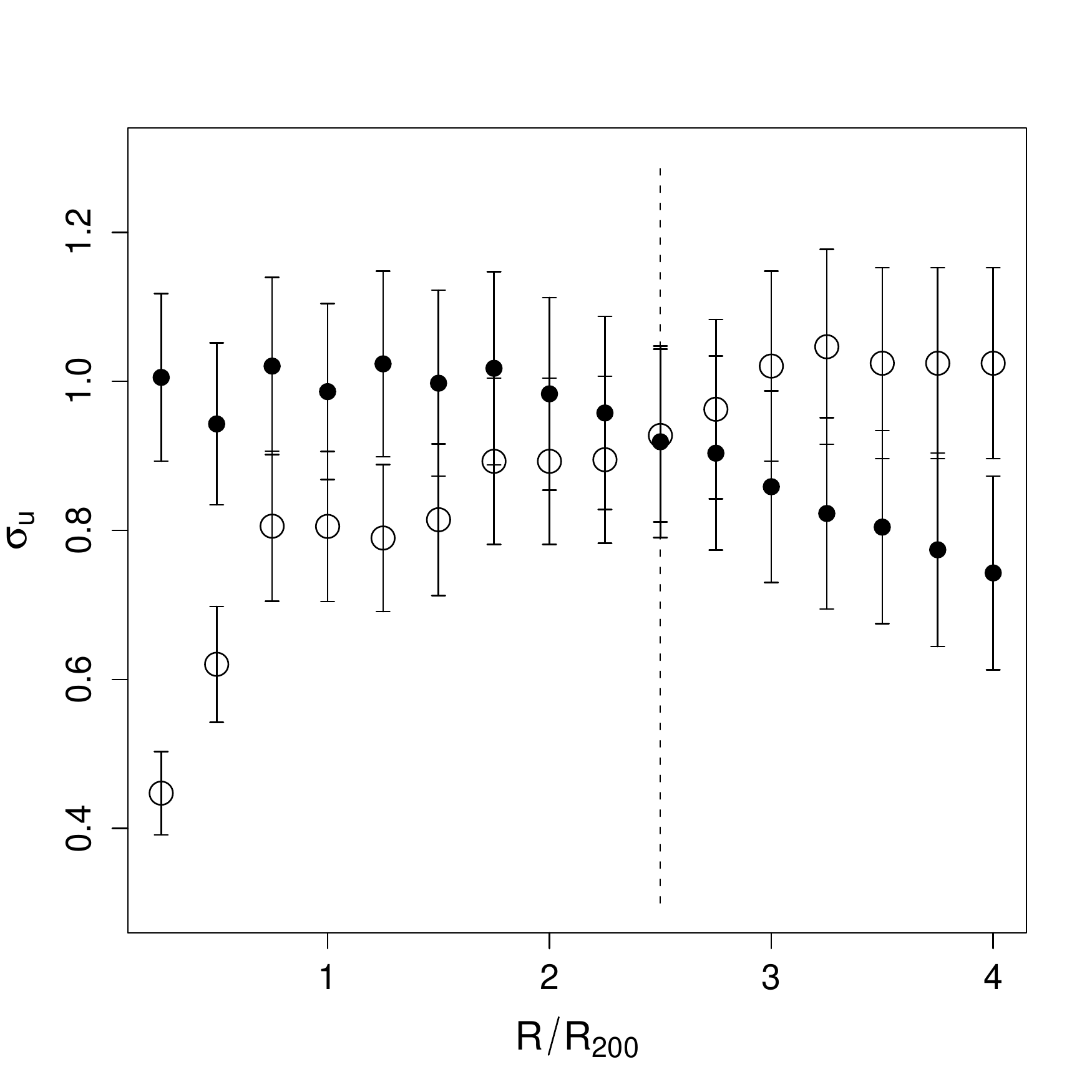}
\caption{Composite groups velocity dispersion as a function of the normalized
radial distances to the group centers. Filled circles denote galaxies in G groups, while open circles denote galaxies in NG groups.}
\label{}
\end{figure}

\begin{figure}
\includegraphics[width=84mm]{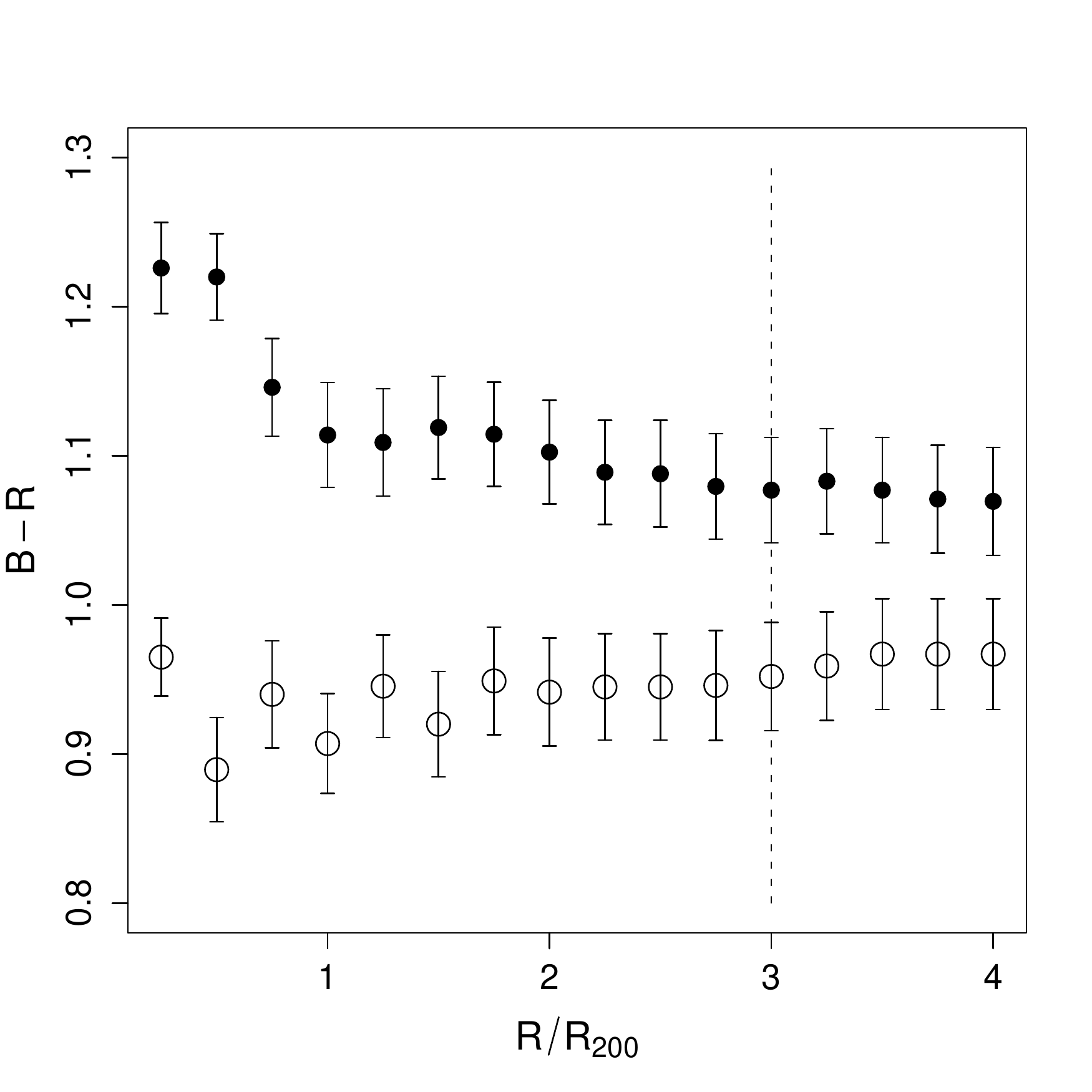}
\caption{B-R color of galaxies in the composite groups as a function of the normalized
radial distances to the group centers. Filled circles denote galaxies in G groups, while open circles denote galaxies in NG groups.}
\label{}
\end{figure}

\begin{figure}
\includegraphics[width=84mm]{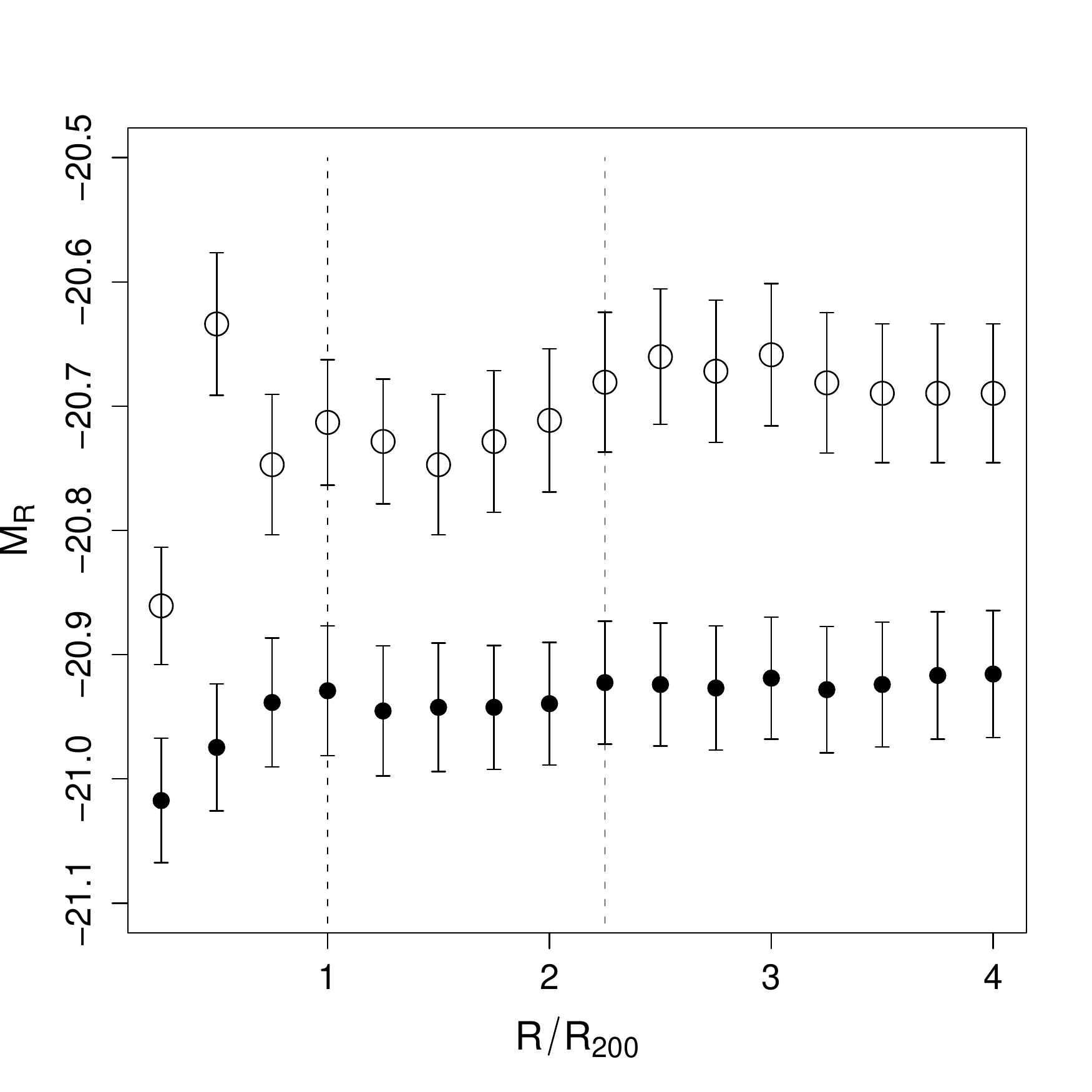}
\caption{Absolute magnitude in R band as a function of the normalized
radial distances to the group centers. Filled circles denote galaxies in G groups, while open circles denote galaxies in NG groups.}
\label{}
\end{figure}

\section{Discussion}

We have studied segregation effects in 57 galaxy groups selected from the 2PIGG catalog (Eke et al. 2004) using 2dF data out to $4R_{200}$. This means we probe galaxy distribution near to the
turnaround radius, thus probably taking into account all members in the infall pattern around the groups
(e.g. Rines \& Diaferio 2006; Cupani, Mezzetti \& Mardirossian 2008).
Instead of focusing our analysis on choosing specific galaxy types to study
segregation, we have used the dynamical state of galaxy systems to test for different
levels of environmental influence on galaxies. The theoretical expectation
is that the underlying velocity distribution is normal for systems in dynamical equilibrium.
Using the AD test, we divided the sample in Gaussian and non-Gaussian groups. These were used
to build the composite G and NG groups. Some general results we found were expected: 
segregation in velocity space (galaxies brighter than $M_R=-21.5$  are moving more slowly than other group galaxies); 
and color and luminosity gradients towards the center of the groups.
However, important differences emerge when we compare the behaviour of galaxies in G and NG groups.
For instance, color gradient and overall reddening are stronger in the case of the G group out to large distances,
showing a significant raise of more evolved galaxies from non-relaxed to relaxed systems (Figure 3).
This is consistent with the luminosity profiles, indicating that galaxies in the G group are
significantly brighter than those in the NG group (Figure 4).
On the other hand, the rising velocity dispersion profile for galaxies in the NG group indicate that, though less
evolved now, galaxies in non-Gaussian systems may be undergoing a more intense phase of interactions 
in their inner parts at the present epoch (Figure 2). These results 
are in agreement with the work of Popesso et al. (2007), in which Abell clusters with an abnormally low X-ray luminosity for their
mass have a higher fraction of blue galaxies, and
are characterized by leptokurtic (more centrally concentrated than a Gaussian) velocity distribution of their member
galaxies in the outskirts ($1.5 < R/R_{200} \leq 3.5$), as expected for systems undergoing a phase of mass accretion.
This also fairly agrees with Osmond \& Ponman (2004) who have found that groups with an abnormally low velocity dispersion
relative to their X-ray properties have a higher fraction of
spirals and could be dynamically unrelaxed. The low velocity dispersions are probably consequence of the
interactions in the inner part of the groups. Since they only considered the central group regions and the
brightest galaxies, our analysis suggests that they may have found unrelaxed systems,
but could have understimated the global velocity dispersions of
these groups (see Table 1 and Figure 2). 

Taken together, these facts point out a scenario where young systems have galaxies bluer and fainter up to large radii ($\sim 4R_{200}$),
possessing lower velocity dispersions in the inner parts (and higher velocity dispersions in the
outer parts) in comparison to more evolved systems.
This latter result is also related to the segregation detected in the velocity space.
Galaxies brighter than $M_R=-21.5$  are moving more slowly than other group galaxies,
but the relation $\sigma_u-M_R$ is steeper for non-Gaussian groups, indicating a departure
from the energy equipartition expectation -- $\sigma_u \propto 10^{0.2M_R}$ (see Figure 1). 
Our work suggests that  the slope of the relation $\sigma_u-M_R$ could be 
used to determine the evolutionary
stage of galaxy groups.

\section*{Acknowledgments}
We thank the referee for very useful suggestions.
We also thank S. Rembold for interesting discussions.
ALBR thanks the support of CNPq, grants 201322/2007-2 and 471254/2008-8. 
PAAL thanks the support of FAPERJ, process 110.237/2010. MT thanks the support of FAPESP, process 2008/50198-3.

\bsp

\label{lastpage}

\end{document}